\catcode`\@=11

\font\tenmsa=msam10
\font\sevenmsa=msam7
\font\fivemsa=msam5
\font\tenmsb=msbm10
\font\sevenmsb=msbm7
\font\fivemsb=msbm5
\newfam\msafam
\newfam\msbfam
\textfont\msafam=\tenmsa  \scriptfont\msafam=\sevenmsa
  \scriptscriptfont\msafam=\fivemsa
\textfont\msbfam=\tenmsb  \scriptfont\msbfam=\sevenmsb
  \scriptscriptfont\msbfam=\fivemsb

\def\hexnumber@#1{\ifnum#1<10 \number#1\else
 \ifnum#1=10 A\else\ifnum#1=11 B\else\ifnum#1=12 C\else
 \ifnum#1=13 D\else\ifnum#1=14 E\else\ifnum#1=15 F\fi\fi\fi\fi\fi\fi\fi}

\def\msa@{\hexnumber@\msafam}
\def\msb@{\hexnumber@\msbfam}
\mathchardef\boxdot="2\msa@00
\mathchardef\boxplus="2\msa@01
\mathchardef\boxtimes="2\msa@02
\mathchardef\square="0\msa@03
\mathchardef\blacksquare="0\msa@04
\mathchardef\centerdot="2\msa@05
\mathchardef\lozenge="0\msa@06
\mathchardef\blacklozenge="0\msa@07
\mathchardef\circlearrowright="3\msa@08
\mathchardef\circlearrowleft="3\msa@09
\mathchardef\rightleftharpoons="3\msa@0A
\mathchardef\leftrightharpoons="3\msa@0B
\mathchardef\boxminus="2\msa@0C
\mathchardef\Vdash="3\msa@0D
\mathchardef\Vvdash="3\msa@0E
\mathchardef\vDash="3\msa@0F
\mathchardef\twoheadrightarrow="3\msa@10
\mathchardef\twoheadleftarrow="3\msa@11
\mathchardef\leftleftarrows="3\msa@12
\mathchardef\rightrightarrows="3\msa@13
\mathchardef\upuparrows="3\msa@14
\mathchardef\downdownarrows="3\msa@15
\mathchardef\upharpoonright="3\msa@16

\mathchardef\downharpoonright="3\msa@17
\mathchardef\upharpoonleft="3\msa@18
\mathchardef\downharpoonleft="3\msa@19
\mathchardef\rightarrowtail="3\msa@1A
\mathchardef\leftarrowtail="3\msa@1B
\mathchardef\leftrightarrows="3\msa@1C
\mathchardef\rightleftarrows="3\msa@1D
\mathchardef\Lsh="3\msa@1E
\mathchardef\Rsh="3\msa@1F
\mathchardef\rightsquigarrow="3\msa@20
\mathchardef\leftrightsquigarrow="3\msa@21
\mathchardef\looparrowleft="3\msa@22
\mathchardef\looparrowright="3\msa@23
\mathchardef\circeq="3\msa@24
\mathchardef\succsim="3\msa@25
\mathchardef\gtrsim="3\msa@26
\mathchardef\gtrapprox="3\msa@27
\mathchardef\multimap="3\msa@28
\mathchardef\therefore="3\msa@29
\mathchardef\because="3\msa@2A
\mathchardef\doteqdot="3\msa@2B

\mathchardef\traceiangleq="3\msa@2C
\mathchardef\precsim="3\msa@2D
\mathchardef\lesssim="3\msa@2E
\mathchardef\lessapprox="3\msa@2F
\mathchardef\eqslantless="3\msa@30
\mathchardef\eqslantgtr="3\msa@31
\mathchardef\curlyeqprec="3\msa@32
\mathchardef\curlyeqsucc="3\msa@33
\mathchardef\preccurlyeq="3\msa@34
\mathchardef\leqq="3\msa@35
\mathchardef\leqslant="3\msa@36
\mathchardef\lessgtr="3\msa@37
\mathchardef\backprime="0\msa@38
\mathchardef\risingdotseq="3\msa@3A
\mathchardef\fallingdotseq="3\msa@3B
\mathchardef\succcurlyeq="3\msa@3C
\mathchardef\geqq="3\msa@3D
\mathchardef\geqslant="3\msa@3E
\mathchardef\gtrless="3\msa@3F
\mathchardef\sqsubset="3\msa@40
\mathchardef\sqsupset="3\msa@41
\mathchardef\trianglerighteq="3\msa@44
\mathchardef\trianglelefteq="3\msa@45
\mathchardef\bigstar="0\msa@46
\mathchardef\between="3\msa@47
\mathchardef\blacktriangledown="0\msa@48
\mathchardef\blacktriangleright="3\msa@49
\mathchardef\blacktriangleleft="3\msa@4A
\mathchardef\blacktriangle="0\msa@4E
\mathchardef\triangledown="0\msa@4F
\mathchardef\eqcirc="3\msa@50
\mathchardef\lesseqgtr="3\msa@51
\mathchardef\gtreqless="3\msa@52
\mathchardef\lesseqqgtr="3\msa@53
\mathchardef\gtreqqless="3\msa@54
\mathchardef\Rrightarrow="3\msa@56
\mathchardef\Lleftarrow="3\msa@57
\mathchardef\veebar="2\msa@59
\mathchardef\barwedge="2\msa@5A
\mathchardef\doublebarwedge="2\msa@5B
\mathchardef\angle="0\msa@5C
\mathchardef\measuredangle="0\msa@5D
\mathchardef\sphericalangle="0\msa@5E
\mathchardef\varpropto="3\msa@5F
\mathchardef\smallsmile="3\msa@60
\mathchardef\smallfrown="3\msa@61
\mathchardef\Subset="3\msa@62
\mathchardef\Supset="3\msa@63
\mathchardef\Cup="2\msa@64

\mathchardef\Cap="2\msa@65

\mathchardef\curlywedge="2\msa@66
\mathchardef\curlyvee="2\msa@67
\mathchardef\leftthreetimes="2\msa@68
\mathchardef\rightthreetimes="2\msa@69
\mathchardef\subseteqq="3\msa@6A
\mathchardef\supseteqq="3\msa@6B
\mathchardef\bumpeq="3\msa@6C
\mathchardef\Bumpeq="3\msa@6D
\mathchardef\lll="3\msa@6E

\mathchardef\ggg="3\msa@6F

\mathchardef\circledS="0\msa@73
\mathchardef\pitchfork="3\msa@74
\mathchardef\dotplus="2\msa@75
\mathchardef\backsim="3\msa@76
\mathchardef\backsimeq="3\msa@77
\mathchardef\complement="0\msa@7B
\mathchardef\intercal="2\msa@7C
\mathchardef\circledcirc="2\msa@7D
\mathchardef\circledast="2\msa@7E
\mathchardef\circleddash="2\msa@7F
\def\ulcorner{\delimiter"4\msa@70\msa@70 }
\def\urcorner{\delimiter"5\msa@71\msa@71 }
\def\llcorner{\delimiter"4\msa@78\msa@78 }
\def\lrcorner{\delimiter"5\msa@79\msa@79 }
\def\yen{\mathhexbox\msa@55 }
\def\checkmark{\mathhexbox\msa@58 }
\def\circledR{\mathhexbox\msa@72 }
\def\maltese{\mathhexbox\msa@7A }
\mathchardef\lvertneqq="3\msb@00
\mathchardef\gvertneqq="3\msb@01
\mathchardef\nleq="3\msb@02
\mathchardef\ngeq="3\msb@03
\mathchardef\nless="3\msb@04
\mathchardef\ngtr="3\msb@05
\mathchardef\nprec="3\msb@06
\mathchardef\nsucc="3\msb@07
\mathchardef\lneqq="3\msb@08
\mathchardef\gneqq="3\msb@09
\mathchardef\nleqslant="3\msb@0A
\mathchardef\ngeqslant="3\msb@0B
\mathchardef\lneq="3\msb@0C
\mathchardef\gneq="3\msb@0D
\mathchardef\npreceq="3\msb@0E
\mathchardef\nsucceq="3\msb@0F
\mathchardef\precnsim="3\msb@10
\mathchardef\succnsim="3\msb@11
\mathchardef\lnsim="3\msb@12
\mathchardef\gnsim="3\msb@13
\mathchardef\nleqq="3\msb@14
\mathchardef\ngeqq="3\msb@15
\mathchardef\precneqq="3\msb@16
\mathchardef\succneqq="3\msb@17
\mathchardef\precnapprox="3\msb@18
\mathchardef\succnapprox="3\msb@19
\mathchardef\lnapprox="3\msb@1A
\mathchardef\gnapprox="3\msb@1B
\mathchardef\nsim="3\msb@1C
\mathchardef\napprox="3\msb@1D
\mathchardef\nsubseteqq="3\msb@22
\mathchardef\nsupseteqq="3\msb@23
\mathchardef\subsetneqq="3\msb@24
\mathchardef\supsetneqq="3\msb@25
\mathchardef\subsetneq="3\msb@28
\mathchardef\supsetneq="3\msb@29
\mathchardef\nsubseteq="3\msb@2A
\mathchardef\nsupseteq="3\msb@2B
\mathchardef\nparallel="3\msb@2C
\mathchardef\nmid="3\msb@2D
\mathchardef\nshortmid="3\msb@2E
\mathchardef\nshortparallel="3\msb@2F
\mathchardef\nvdash="3\msb@30
\mathchardef\nVdash="3\msb@31
\mathchardef\nvDash="3\msb@32
\mathchardef\nVDash="3\msb@33
\mathchardef\ntrianglerighteq="3\msb@34
\mathchardef\ntrianglelefteq="3\msb@35
\mathchardef\ntriangleleft="3\msb@36
\mathchardef\ntriangleright="3\msb@37
\mathchardef\nleftarrow="3\msb@38
\mathchardef\nrightarrow="3\msb@39
\mathchardef\nLeftarrow="3\msb@3A
\mathchardef\nRightarrow="3\msb@3B
\mathchardef\nLeftrightarrow="3\msb@3C
\mathchardef\nleftrightarrow="3\msb@3D
\mathchardef\divideontimes="2\msb@3E
\mathchardef\varnothing="0\msb@3F
\mathchardef\nexists="0\msb@40
\mathchardef\mho="0\msb@66
\mathchardef\thorn="0\msb@67
\mathchardef\beth="0\msb@69
\mathchardef\gimel="0\msb@6A
\mathchardef\daleth="0\msb@6B
\mathchardef\lessdot="3\msb@6C
\mathchardef\gtrdot="3\msb@6D
\mathchardef\ltimes="2\msb@6E
\mathchardef\rtimes="2\msb@6F
\mathchardef\shortmid="3\msb@70
\mathchardef\shortparallel="3\msb@71
\mathchardef\smallsetminus="2\msb@72
\mathchardef\thicksim="3\msb@73
\mathchardef\thickapprox="3\msb@74
\mathchardef\approxeq="3\msb@75
\mathchardef\succapprox="3\msb@76
\mathchardef\precapprox="3\msb@77
\mathchardef\curvearrowleft="3\msb@78
\mathchardef\curvearrowright="3\msb@79
\mathchardef\digamma="0\msb@7A
\mathchardef\varkappa="0\msb@7B
\mathchardef\hslash="0\msb@7D
\mathchardef\hbar="0\msb@7E
\mathchardef\backepsilon="3\msb@7F
\def\Bbb{\ifmmode\let\next\Bbb@\else
 \def\next{\errmessage{Use \string\Bbb\space only in math mode}}\fi\next}
\def\Bbb@#1{{\Bbb@@{#1}}}
\def\Bbb@@#1{\fam\msbfam#1}

\catcode`\@=\active

\def\CM{\hbox{{$\cal M$}}}

\def\R{{\Bbb R}}
\def\C{{\Bbb C}}
\def\Z{{\Bbb Z}}

\def\vect{{\bf t}}\def\vecv{{\bf v}}
\def\vecu{{\bf u}}\def\vecx{{\bf x}}
\def\veca{{\bf a}}

\def\<{\langle}
\def\>{\rangle}
\def\del{{\partial}}
\def\lform{\hbox{$\sqcup$}\llap{\hbox{$\sqcap$}}}

\def\eps{{\epsilon}}

\def\trace{{\rm Tr\, }}

\def\dcross{{\bowtie}}

\def\cosub{{\Delta\kern -.65em \raisebox{.02em}{-}\kern .35em}}

\def\tens{\mathop{\otimes}}

\def\span{{\rm span}}

\def\id{{\rm id}}

\def\Vhaj{{V\haj{\ }}}
\def\proof{\goodbreak\noindent{\bf Proof\quad}}

\def\endproof{{\ $\lform$}\bigskip }

\def\und#1{{\underline {#1}}}
\def\haj#1{{\mathaccent20 {#1}}}

\def\note#1{}

\def\equad{\kern -1.7em}

\def\eqn#1#2{\begin{equation}#2\label{#1}\end{equation}}

\def\ceqn#1#2{\begin{equation}\label{#1}\begin{array}{c}#2
\end{array}\end{equation}}
\def\align#1{\begin{eqnarray*}#1\end{eqnarray*}}
\def\alignn#1#2{\begin{eqnarray}\label{#1}#2
\end{eqnarray}}

\def\dila{{\varsigma}} 

\documentstyle[11pt, epsf]{article}
\def\d{{{\rm d}}}

\newtheorem{lemma}{Lemma}[section]
\newtheorem{propos}[lemma]{Proposition}

\begin{document}\baselineskip 23pt

{\ }\hskip 4.7in DAMTP/94-20
\vspace{.2in}

\begin{center} {\LARGE $q$-EPSILON TENSOR FOR QUANTUM AND\\ BRAIDED SPACES}
\\ \baselineskip 13pt{\ }
{\ }\\ S. Majid\footnote{Royal Society University Research Fellow and Fellow of
Pembroke College, Cambridge}\\
{\ }\\
Department of Applied Mathematics \& Theoretical Physics\\
University of Cambridge, Cambridge CB3 9EW
\end{center}

\begin{center}
February, 1994 -- revised July 1994\end{center}
\vspace{10pt}
\begin{quote}\baselineskip 13pt
\noindent{\bf Abstract}
The machinery of braided geometry introduced previously is
used now to construct the $\epsilon$ `totally antisymmetric tensor' on a
general braided vector space determined by R-matrices. This includes natural
$q$-Euclidean and $q$-Minkowski spaces. The formalism is completely covariant
under the corresponding quantum group such as $\widetilde{SO_q(4)}$ or
$\widetilde{SO_q(1,3)}$. The Hodge $*$ operator and differentials are also
constructed in this approach.
\end{quote}
\baselineskip 23pt

\section{Introduction}

In this paper we apply the systematic theory of braided geometry introduced
during the last few years by the
author\cite{Ma:exa}\cite{Ma:lin}\cite{Ma:poi}\cite{Ma:fre}\cite{Ma:add} to the
problem of defining the totally antisymmetric tensor $\eps_{ijkl}$ and other
antisymmetrisers on quantum spaces of R-matrix type, for the first time in a
general way.

Braided-geometry differs from other approaches to q-deforming physics in that
the deformation is put directly into non-commutativity or `braid statistics' of
the tensor product of independent systems. Individual algebras also tend to be
non-commutative (as in non-commutative geometry) but this is a secon\d ary
phenomenon. The theory is modelled on ideas of super-geometry with a braiding
$\Psi$ (typically defined by a parameter $q$) in the role of $\pm 1$ for usual
bose or fermi statistics. It turns out that this point of view is rather
powerful and using it a great many problems encountered in other approaches
are immediately overcome.

The starting point of braided geometry is that quantum group covariance, unlike
usual group covariance, induces braid statistics\cite{Ma:exa}\cite{Ma:lin}. The
quantum group plays the role of $\Z_2$-grading in the theory of super-symmetry
even when the quantum group is very far from discrete (e.g. when it is the
$q$-deformed Lorentz group). The systematic development of braided geometry has
been a matter of going back to basics and re-inventing from scratch the most
fun\d amental concepts in physics on this basis, layer by layer. After
covariance, the next layer is that of coaddition on quantum spaces, introduced
in \cite{Ma:poi}. Once one can add vectors on braided or q-deformed vector
spaces, the next layer is differentiation, introduced in \cite{Ma:fre} as an
infinitesimal coaddition:
\eqn{diff}{ \del^i f(\vecx)={\rm coeff\ of\ }a_i\ {\rm in}\ f(\veca+\vecx)}
where the addition is a braided addition (so $\veca$ and $\vecx$
braided-commute). Following this, there is also translation-invariant
integration\cite{KemMa:alg}. Braided matrices, traces etc were also introduced
in
\cite{Ma:exa}\cite{Ma:lin}. The approach also links up with the more usual
approach based on quantum forms and non-commutative geometry by pushing the
arguments of \cite{WesZum:cov} backwards (from partial derivatives $\del^i$ to
exterior derivative $\d$). This is essentially known though some details are
included in the present paper for completeness. It provides a constructive
approach
to $\d$.

The antisymmetric tensor  by contrast needs a conceptually new point of view in
order to be able to apply this existing braided geometry. Here we present a
novel and, we believe, powerful point of view about it. This point of view
is useful even when $q=1$ where it corresponds to the view  that the exterior
algebra of forms can and should be viewed as a super-space with co-ordinates
$\theta_i$, say. Usually, one
realises super-spaces using exterior algebras, but our point of view is the
reverse of this. The braided geometry applies just as well to super-spaces and
their q-deformations as to bosonic-spaces and their deformations, so we can
apply it at once to the exterior algebra without effort. In particular  it is
natural for us to define
\eqn{eps}{\eps^{i_1i_2\cdots i_n}={\del\over\del \theta_{i_1}}{\del\over\del
\theta_{i_2}}\cdots{\del\over\del \theta_{i_n}} \theta_1\theta_2\cdots
\theta_n}
on any reasonable n-dimensional braided space with top form
$\theta_1\theta_2\cdots\theta_n$. We also construct the Hodge $*$-operator
and interior products on forms in this setting.

Finally, important examples such as $q$-Euclidean and $q$-Minkowski spaces are
also known in this framework of braided geometry\cite{Mey:new}\cite{Ma:euc},
which
examples are compatible too with the earlier ideas of
\cite{CWSSW:lor}\cite{OSWZ:def}\cite{PodWor:def} based on spinors. Hence our
results apply at
once to these important braided spaces.

During the preparation of this paper there appeared \cite{Mey:wav} in which the
$q$-epsilon tensor in the case of
$q$-Minkowski space was found directly by computer and used to develop Hodge
theory and
scaler electrodynamics. Our general formulation in Section~3 is motivated in
part by
this. We would also like to mention \cite{Fio:det} where $q$-epsilon tensors
for
$SO_q(n)$-covariant Euclidean spaces were considered, again rather explicitly.
The tensor
for $GL_q$-covariant quantum planes is even more well known. By contrast with
such specific examples, we present here a uniform R-matrix approach.

\subsection*{Acknowledgements} This work was completed during a visit to the
Erwin Schr\"odinger Institute, Wien. I would like to thank the staff there for
their
help and support.

\subsection*{Preliminaries on braided vector spaces}

Here we recall the formulation in \cite{Ma:poi} of braided vector and covector
spaces, and strengthen their construction slightly for our purposes. The
position
co-ordinates $\vecx=\{x_i\}$ form a braided-covector space, while their
differentials $\del^i$ form a braided vector space\cite{Ma:fre}. Throughout
this paper, we treat only spaces of this type, i.e. braided versions of $\R^n$.

The input \d ata for these constructions are a pair of R-matrices $R,R'\in
M_n\tens M_n$ such that\cite{Ma:poi}
\[ R'_{12}R_{13}R_{23}=R_{23}R_{13}R'_{12},\quad
R'_{23}R_{13}R_{12}=R_{12}R_{13}R'_{23} \]
\[ R_{12}R_{13}R_{23}=R_{23}R_{13}R_{12},\quad (PR+1)(PR-1)=0,\quad
R_{21}R'=R'_{21}R\]
where $P$ is the usual permutation matrix. These are enough to ensure that
there are braided vector and covector spaces
\[ \Vhaj(R',R)=\{x_i\}:\ \vecx_1\vecx_2=\vecx_2\vecx_1R',\quad
V(R',R)=\{v^i\}:\ \vecv_1\vecv_2=R'\vecv_2\vecv_1\]
with braided coaddition $\vecx''=\vecx+\vecx'$, $\vecv''=\vecv+\vecv'$ where
these obey the same relations provided $\vecx,\vecv$ and their identical primed
copies have braid statistics
\[ \vecx'_1\vecx_2=\vecx_2\vecx_1R,\quad \vecv'_1\vecv_2=R\vecv_2\vecv'_1.\]
There are also braid statistics between $\vecx$ and $\vecv$ etc.
Mathematically, they form braided-Hopf algebras in the braided category of
$A(R)$-comodules where $A(R)$ is the usual quantum group or bialgebra
associated to $R$. For regular R-matrices they also live in the braided
category of $\widetilde{A}$-comodules, where $A$ is a Hopf algebra quotient of
$A(R)$ and $\widetilde A$ is dilatonic extension\cite{Ma:poi}. The
transformation laws are $\vecx\to \vecx\vect\dila$ and
$\vecv\to\dila^{-1}\vect^{-1}\vecv$ where $\vect$ is the quantum matrix
generator and $\dila$ the dilaton.

To this framework, we now add the additional conditions
\[ R_{12}R'_{13}R'_{23}=R'_{23}R'_{13}R_{12},\quad
R_{23}R'_{13}R'_{12}=R'_{12}R'_{13}R_{23} \]
\[ R'_{12}R'_{13}R'_{23}=R'_{23}R'_{13}R'_{12}\]
so that there is a certain symmetry between $R$ and $R'$. More precisely, we
have a symmetry
\[ R \leftrightarrow -R'\]
and can thereby define
\[ \Lambda(R',R)\equiv \Vhaj(-R,-R')=\{\theta_i\},\quad
\Lambda^*(R',R)\equiv V(-R,-R')=\{\phi^i\}\]
which we call respectively the braided covector space of antisymmetric tensors
or {\em forms} $\Lambda$ and braided vector space of {\em coforms} $\Lambda^*$.
As a braided-Hopf algebra, the
latter is the dual of the former. In our geometrical application, the
differentials $\theta_i=\d x_i$ obey the algebra of forms, while the operators
$\del\over\del\theta_i$ necessarily obey the algebra of coforms.
The forms and coforms are covariant under the transformation
$\theta\to\theta\vect\dila$ and $\phi\to\dila^{-1}\vect^{-1}\phi$ of a quantum
group obtained from $A(-R')=A(R')$. We assume for convenience that this is the
same as the quantum group obtained from $A(R)$. This is true in some
generality, for
example whenever $PR'=f(PR)$ for some function $f$. It is also true for our
$q$-Euclidean
and $q$-Minkowski examples. See \cite{Mey:new} for the latter.

\section{Symmetric and antisymmetric tensors by differentiation}

In \cite{Ma:fre} was introduced a general theory of partial differentiation
$\del^i$ on braided spaces of the type above. This recovered all known cases
and, moreover, works generally. If $\{x_i\}$ are the position co-ordinates,
then $\del^i$ are given explicitly as the operators\cite{Ma:fre}
\eqn{delx}{ {\del\over\del x_i}(\vecx_1\cdots \vecx_m)={\bf
e}^i{}_1\vecx_2\cdots\vecx_m \left[m;R\right]_{1\cdots m}}
where ${\bf e}^i$ is a basis covector $({\bf e}^i){}_j=\delta^i{}_j$, i.e.
\[ {\del\over\del x_i}(x_{i_1}\cdots x_{i_m})=\delta^i{}_{j_1}x_{j_2}\cdots
x_{j_m}\left[m;R\right]^{j_1\cdots j_m}_{i_1\cdots i_m}.\]
Here
\eqn{braint}{\left[m;R\right]=1+(PR)_{12}+(PR)_{12}(PR)_{23}
+\cdots+(PR)_{12}\cdots (PR)_{m-1,m}}
are the {\em braided integers} which we introduced for this purpose. One of the
main theorems in \cite{Ma:fre} is that these differentiation operators on
$\{x_i\}$ obey the vector relations as for the $\{v^i\}$. One can say that the
partial-derivatives  $R'$-commute. They also obey a braided-Leibniz rule with
braiding  $R$\cite{Ma:fre}.

Moreover, since the result is quite general, it holds just as well for the
partial derivatives $\del\over\del\theta_i$,
\eqn{deltheta}{{\del\over\del \theta_i}(\theta_1\cdots \theta_m)={\bf
e}^i{}_1\theta_2\cdots\theta_m \left[m;-R'\right]_{1\cdots m}}
on the algebra of forms $\{\theta_i\}$. We deduce that these differential
operators obey the relations of the coforms $\Lambda^*$. This means that they
$-R$-commute and obey a braided Leibniz rule with braiding $-R'$.

These theorems about the partial-derivatives are quite powerful, and we use
them now. In particular we can differentiate any function $f$ and will know
that
\[ v^{i_1i_2\cdots i_m}={\del\over\del x_{i_1}}\cdots {\del\over\del
x_{i_m}}f\]
is an $R'$-symmetric tensor of rank $m$, in the sense
\eqn{R'-sym}{ R'^{i_p i_{p+1}}_{ab} v^{i_1\cdots i_{p-1}bai_{p+2}\cdots
i_m}=v^{i_1\cdots i_m},\quad \forall p=1,\cdots,m-1.}
If $f$ is a scaler function (quantum group covariant) then, because all our
constructions in \cite{Ma:fre} are covariant, we will know that this tensor is
likewise invariant. The same applies in the $\theta$ space, in which case the
tensors must be manifestly $-R$-symmetric i.e., $R$-antisymmetric in the sense
\eqn{R-asym}{ R^{i_p i_{p+1}}_{ab} \phi^{i_1\cdots i_{p-1}bai_{p+2}\cdots
i_m}=-\phi^{i_1\cdots i_m},\quad \forall p=1,\cdots,m-1.}

For a simple example of this idea, we suppose that the co-ordinate algebra
$\{x_i\}$ has a {\em radius function} $r^2$ which is quantum-group invariant (a
scaler under the transformation). Then
\[ \eta^{ij}={\del\over\del x_i}{\del\over\del x_j} r^2\]
is an $R'$-symmetric invariant tensor, which we call the {\em metric}
associated to the radius function. If $r^2$ is quadratic then $\eta$ is an
ordinary $\C$-number matrix. In nice cases it will be invertible. Moreover,
invariance implies at once the first half of the identities
\eqn{ggR}{ \eta_{ia}\eta_{jb}R^a{}_m{}^b{}_n=R^a{}_i{}^b{}_j
\eta_{am}\eta_{bn}, \quad \eta_{ia}\eta_{jb}R'^a{}_m{}^b{}_n=R'^a{}_i{}^b{}_j
\eta_{am}\eta_{bn}}
We adopt the second half too in order to keep the symmetry between $R$ and
$-R'$. They have the meaning that then\cite{Mey:new} the algebra of vectors and
covectors are isomorphic via
\[ x_i=\eta_{ia}v^a,\quad v^i=x_a\eta^{ai},\quad
\eta_{ja}\eta^{ia}=\delta^i{}_j=\eta_{aj}\eta^{ai}\]
so that the metric can be used to raise and lower indices for any operators
behaving like the vectors and covectors. It clearly does the same job for
raising and lowering indices of the forms and coforms by the symmetry.

We now use the same idea in the deformed super-space of forms. We say that the
braided space has {\em form dimension} $n$ if the algebra of forms has (up to
normalisation) a unique element of highest degree $n$, which we call the {\em
top form} $\omega$. In nice cases the form dimension will be the same as the
number $n$ of our co-ordinate generators and indeed, the top form
will be $\omega=\theta_1\cdots\theta_n$.
We then define
\[ \eps^{i_1 i_2\cdots i_n}={\del\over\del \theta_{i_1}}\cdots {\del\over\del
\theta_{i_n}} \omega={\del\over\del \theta_{i_1}}\cdots {\del\over\del
\theta_{i_n}}\theta_1\cdots\theta_n.\]
By the reasoning above, it will be $R$-antisymmetric.

If we want to have tensors with lower indices, we can obtain them also by
differentiation of monomials in the co-ordinates. Thus an $R'$-symmetric tensor
of rank $m$ with lower indices means
\eqn{symlower}{ x_{i_1\cdots i_{p-1}bai_{p+2}\cdots
i_m}R'{}^a{}_{i_p}{}^b{}_{i_{p+1}}=x_{i_1\cdots i_m},\quad \forall
p=1,\cdots,m-1}
and an $R$-antisymmetric tensor with lower indices means
\eqn{asymlower}{ \theta_{i_1\cdots i_{p-1}bai_{p+2}\cdots
i_m}R{}^a{}_{i_p}{}^b{}_{i_{p+1}}=-\theta_{i_1\cdots i_m},\quad \forall
p=1,\cdots,m-1.}
The first of these can be obtained by applying any m-th order differential
operator built from $\del\over\del x_i$ to monomials $x_{i_1}\cdots x_{i_m}$.
Likewise, we can follow the same idea in form-space and obtain an
$R$-antisymmetric tensor by applying
an $m$-th order operator built from $\del\over\del\theta_i$ to
$\theta_{i_1}\cdots\theta_{i_m}$. For example, we define
\[ \eps_{i_1\cdots i_n}={\del\over\del \theta_n}\cdots {\del\over\del
\theta_1} \theta_{i_1}\cdots\theta_{i_n}.\]
Its total R-antisymmetry is inherited this time from antisymmetry of the
$\theta_i$ co-ordinates in form-space.

\begin{propos} If the top form is $\omega=\theta_1\cdots\theta_n$ say, we have
an explicit formula:
\[ \eps^{i_1\cdots i_n}=([n;-R']!)^{i_n \cdots i_1}_{12\cdots n},\quad
\eps_{i_1\cdots i_n}=([n;-R']!)_{i_1 \cdots i_n}^{12\cdots n}\]
where
\align{ &&\equad [n;-R']!=(\id\tens [2;-R'])(\id\tens [3;-R'])\cdots [n;-R']\\
&&=(1-PR'_{n-1\ n})(1-PR'_{n-2\ n-1}+PR'_{n-2\ n-1}PR'_{n-1\ n})\cdots\\
&&\cdots (1-PR'_{12}+PR'_{12}PR'_{23}-\cdots+(-1)^{n-1}PR'_{12}PR'_{23}\cdots
PR'_{n-1\ n})}
is built from braided-integers (\ref{braint}).
\end{propos}
\proof This follows directly from the above definitions by carefully iterating
the formula (\ref{deltheta})
for braided-differentiation on the $\theta$ co-ordinates. \endproof

For example, in four dimensions, the braided 4-factorial matrix is
\[  ([4;-R']!)^{i_1\cdots i_4}_{j_1\cdots j_4}=[2,-R']^{i_3
i_4}_{b_3b_4}[3,-R']^{i_2 b_3 b_4}_{a_2 a_3 a_4}[4; -R']^{i_1 a_2 a_3
a_4}_{j_1j_2j_2j_4}\]
and is totally $R$-antisymmetric in its lower indices and in its upper-indices.

If one wants totally antisymmetric tensors of lower rank, these are provided by
the lower braided-factorials $[m,-R']!$ in the same way. For example
\[
{\del\over\del\theta_{i_1}}{\del\over\del\theta_{i_2}}\theta_{j_1}\theta_{j_2}
=[2;-R']^{i_2i_1}_{j_1j_2},\quad [2;-R']=1-PR'\]
\[
{\del\over\del\theta_{i_1}}{\del\over\del\theta_{i_2}}
{\del\over\del\theta_{i_3}}
\theta_{j_1}\theta_{j_2}\theta_{j_3}=([3;-R']!)^{i_3i_2i_1}_{j_1j_2j_3},\quad
[3;-R']!=(1-PR'_{23})(1-PR'_{12}+PR'_{12}PR'_{23})\]
etc. One can also take different numbers of $\theta$ derivatives and
co-ordinates, giving tensors with different numbers of lower and upper indices,
but again totally $R$-antisymmetric among the lower and among the upper.

If one wants tensors with totally $R'$-symmetric inputs and outputs, these are
provided by  braided-factorials $[n:R]!$ and other braided-integers, with $R$
in place of $-R'$. The symmetric and antisymmetric theory here are just the
same construction, with a different choice of R-matrix. In this context,
there is already proven a {\em braided-binomial theorem} in \cite{Ma:fre} for
`counting' such `braided permutations'.

\section{Application to Hodge $*$-operator}

One can  obtain still more tensors with symmetric or antisymmetric inputs and
outputs by contraction along the lines
of \cite{Mey:wav}. For example, given our $\eps$ tensors it is natural to
consider the
contractions of $n-m$ indices,
\eqn{proj}{ P_{i_1\cdots i_m}{}^{j_1\cdots j_m}=\eps_{i_1\cdots i_m
a_{m+1}\cdots a_n}\eps^{j_1\cdots j_ma_n\cdots
a_{m+1}}.}
These are typically proportional to projection operators, i.e.
\[ P_{i_1\cdots i_m}{}^{a_n\cdots a_{m+1}}P_{a_{m+1}\cdots a_n}{}^{j_m\cdots
j_1}=d_m P_{i_1\cdots i_m}{}^{j_m\cdots j_1}\]
for some constants $\{d_m\}$. This is verified in examples, where also these
constants can be determined. On the other hand, it appears to be a rather
general feature which can be expected for any suitably nice $R,R'$-matrices.
These $P$ project onto tensors with totally R-antisymmetric indices.

\begin{propos} There is a well-defined operator on forms given by
\[ P:\Lambda\to\Lambda,\quad P(\theta_{i_1}\cdots \theta_{i_m})=P_{i_1\cdots
i_m}{}^{a_m\cdots a_1}\theta_{a_1}\cdots\theta_{a_m}=d_m
\theta_{i_1}\cdots\theta_{i_m}\]
\end{propos}
\proof One can expect the diagonal form in view of the above remarks since the
products $\theta_{i_1}\cdots \theta_{i_m}$ are already $R$-antisymmetric. Here
we check that $P$ as an operator is indeed well-defined. Indeed, the relations
of $\Lambda$ are respected as
\[
P(\theta_{i_1}\cdots\theta_b\theta_a\cdots\theta_{i_m})R^a{}_{i_p}{}^b
{}_{i_{p+1}}=\eps_{i_1\cdots ba\cdots i_m a_{m+1}\cdots a_n}
\eps^{b_1\cdots b_ma_n\cdots
a_{m+1}}\theta_{b_1}\cdots\theta_{b_m}=-P(\theta_{i_1}\cdots\theta_{i_m})\]
for all $p$ due to $R$-antisymmetry of $\eps$. We give this is in detail
because this and a similar consideration for the output of $P$ dictates the
ordering of the indices in the action of $P$. \endproof

As another immediate application of our epsilon tensor one can
write a general R-matrix formula for the quantum-determinant of the
symmetry quantum group of our theory, namely
\[ \det(\vect)=d_0^{-1} \eps_{i_1\cdots i_n}t^{i_1}{}_{j_1}\cdots
t^{i_n}{}_{j_n}\eps^{j_n\cdots j_1}=([n;-R']!)^{1\cdots n}_{i_1\cdots
i_n}t^{i_1}{}_{j_1}\cdots t^{i_n}{}_{j_n}  ([n;-R']!)^{j_1\cdots j_n}_{1\cdots
n}.\]

There is no metric needed here, but if one exists then it is easy to see
from (\ref{ggR}) that it can used to turn any  $R'$-symmetric or
$R$-antisymmetric tensor with upper-indices to one with lower indices.
In our setting with a
unique top form, one can also expect that
a totally antisymmetric tensor with $n$ indices is unique up to a scale. In
this
case one has
\eqn{epsud}{ \eps_{i_1i_2\cdots i_n}=\lambda
\eta_{i_1 a_1}\cdots \eta_{i_n a_n}\eps^{a_1\cdots a_n}=\lambda
\eps^{a_1\cdots a_n}\eta_{a_1 i_1}\cdots \eta_{a_n i_n}}
where $\lambda$ is a constant depending on the metric.

Finally, one can use the $\eps$ tensor in the usual way to define a Hodge
$*$-operator, along the lines in \cite{Mey:wav} for $q$-Minkowski space, where
$\eps$ was found by hand. In the present formulation we have
\begin{propos} There is a well-defined operator on forms given by
\eqn{hodge}{*:\Lambda\to\Lambda,\quad (\theta_{i_1}\cdots
\theta_{i_m})^*=\eps^{a_1\cdots a_m b_n\cdots b_{m+1}}\eta_{a_1
i_1}\cdots\eta_{a_m i_m}\theta_{b_{m+1}}\cdots \theta_{b_n}= H_{i_1\cdots
i_m}{}^{a_n\cdots a_{m+1}}\theta_{a_{m+1}}\cdots \theta_{a_n}.}
\end{propos}
\proof This time consistency with the relations of $\Lambda$ follows using
(\ref{ggR}) after which we can then use antisymmetry of $\eps$ as in the
preceding proposition. \endproof

The appropriate tensor $H$ here has square which is typically proportional to
the projectors in (\ref{proj}),
\[ H_{i_1\cdots i_m}{}^{a_n\cdots a_{m+1}}H_{a_{m+1}\cdots a_n}{}^{j_m\cdots
j_1}\propto P_{i_1\cdots i_m}{}^{j_m\cdots j_1}.\]
This too is verified in examples, were one also learns the constants of
proportionality. It holds in some generality and
means that $*^2\propto\id$ on forms of each degree. This is analogous to the
classical situation. Motivated by this one can also define the interior product
of forms by a form $\theta_i$ as the adjoint under $*$ of multiplication by
$\theta_i$ in the exterior algebra. It obeys a graded $\Z_2$-Leibniz rule, as
checked for $q$-Minkowski case in \cite{Mey:wav}.

It is possible also to make a much more radical formulation of the interior
product and Hodge $*$ operations, based on the idea of differentiation
on form space and not directly on $\eps$. Thus we define the braided-interior
product $i$ and braided-Hodge operator $\circ$ in the algebra of forms
$\{\theta_i\}$ by
\[ i_{f(\theta)} g(\theta)=f(\del)g(\theta),\quad f(\theta)^\circ
=i_{f(\theta)}\theta_1\cdots\theta_n\]
where $f(\del)$ consists of relacing $\theta_i$ by the operators
$\del_i=\eta_{ia}{\del\over\del\theta_a}$. For $\circ$ we use
$\theta_1\cdots\theta_n$ (say) as the top form. We have explicitly,
\alignn{newhodge}{ (\theta_{i_1}\cdots\theta_{i_m})^\circ\equad&&=
\eta_{i_1a_1}\cdots \eta_{i_m a_m}{\del\over\del
\theta_{a_1}}\cdots{\del\over\del\theta_{a_m}}\theta_1\cdots\theta_n\nonumber\\
&&=
\eta_{i_1a_1}\cdots \eta_{i_m a_m}((\id\tens [n-m+1;-R'])\cdots
[n;-R'])^{a_m\cdots a_1 b_{m+1}\cdots b_n}_{12\cdots n}\theta_{b_{m+1}}\cdots
\theta_{b_{n}}.}

For example, in four dimensions, the formulae are
\[
(\theta_{i_1}\theta_{i_2}\theta_{i_3}\theta_{i_4})^\circ
=\lambda^{-1}\eps_{i_1\cdots i_4}\]
\[
(\theta_{i_1}\theta_{i_2}\theta_{i_3})^\circ=\eta_{i_1a_1}
\eta_{i_2a_2}\eta_{i_3a_3}\eps^{a_3a_2a_1b}\theta_b\]
\[
(\theta_{i_1}\theta_{i_2})^\circ=\eta_{i_1a_1}\eta_{i_2a_2}
(\id\tens[3;-R'])[4;-R']^{a_2 a_1 b_1 b_2}_{1234}\theta_{b_1}\theta_{b_2}\]
\[
(\theta_{i})^\circ=\eta_{ia}[4;-R']^{ab_1b_2b_3}_{1234}\theta_{b_1}
\theta_{b_2}\theta_{b_3},\quad 1^\circ=\theta_1\theta_2\theta_3\theta_4.\]
Note that only the Hodge operation on $n$ and $n-1$ degree forms involve the
braided-factorial or $\eps$ tensor directly. The other degrees involving this
and normalisation integers $1\over (n-m)!$ etc. are replaced now by
differentiation.

This second approach to the Hodge operation is different from the first one,
though agrees when $q=1$ after suitable  normalisations at each degree. In
general we do not have that $i$ is a graded derivation  and we also do not have
that  $\circ^2\propto\id$ on forms of a given degree. On the other hand, this
second approach is conceptually quite simple and can be thought of in fact as a
kind of `Fourier transform' in form-space $\{\theta_i\}$. This is suggested by
the interior product $i_f$ appearing as braided-differentiation in form-space.
Moreover, from this point of view one would expect $\circ^2$ to be something
like the braided-antipode $S$ on the braided-Hopf algebra $\{\theta_i\}$, which
is not simply $\pm1$ in the braided case. The technology for braided Fourier
transform is in \cite{KemMa:alg}.

\section{Differential forms}

Until now we have considered the algebra of forms $\theta_i$ in
isolation, as some q-deformed superspace. For completeness we now consider both
the co-ordinates
$\CM=\{x_i\}$ and the forms $\theta_i$ together with $\theta_i=dx_i$. Thus we
consider the {\em exterior algebra}
\eqn{Omega}{\Omega=\Lambda\und\tens \CM,\quad \vecx_1\theta_2=\theta_2\vecx_1R}
where the product is the braided tensor product with the cross-relations as
stated. The essence of the braided tensor product here is that it keeps all
constructions covariant, hence $\Omega$ remains a comodule algebra under our
background quantum group $\widetilde{A}$ (I would like to thank A. Sudbery for
this remark\cite{Sud:ort}).

Next we consider $\Omega$ as bi-graded with components
\[ \Omega^{p|q}=\span\{\theta_{i_1}\cdots\theta_{i_p} x_{j_1}\cdots
x_{j_q}\},\quad \Omega^p=\oplus_{q=0}^{\infty}\Omega^{p|q},\quad
\Omega^{|q}=\oplus_{p=0}^{\infty}\Omega^{p|q}\]
where $\Omega^p$ are the usual $p$-forms in differential geometry. Actually,
one can proceed quite symmetrically with $\Omega^{|q}$ the `differential forms
in super-space' generated by $x_i=d\theta_i$.

\begin{propos} We define the exterior derivative $\d$ as
\[ \d: \Omega^p\to\Omega^{p+1},\quad d (\theta_{i_1}\cdots\theta_{i_p}
f(\vecx))=\theta_{i_1}\cdots\theta_{i_p} \theta_i {\del\over\del x_i}
f(\vecx).\]
It obeys a right-handed $\Z_2$-graded-Leibniz rule
\[ \d(fg)=(-1)^p(\d f)g+f\d g,\quad\forall g\in\Omega,\ f\in\Omega^p.\]
\end{propos}
\begin{figure}
\begin{picture}(200,280)(-50,0)
 \epsfbox{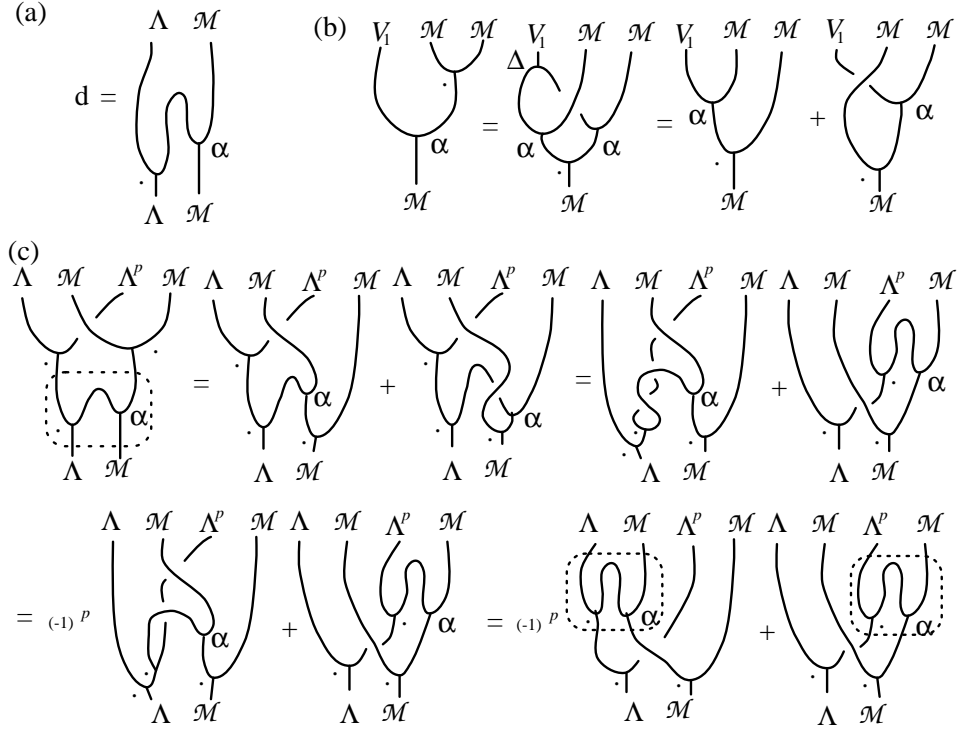}
\end{picture}
\caption{(a) definition of exterior differential $\d$ in braided approach (b)
braided-Leibniz rule for action of partial derivatives and (c) proof of usual
graded-Leibniz rule for $\d$}
\end{figure}
\proof This is well-defined because the partial derivatives $\del\over\del x_i$
are
well-defined as operators on $\CM=V^*$\cite{Ma:fre}. I.e. our $\d$ is built up
from
well-defined operations. It is also covariant under the quantum group
$\widetilde{A}$
for the same reasons. Here the element 
in
$\Lambda\tens V$ (where $V$ is the
vector algebra) is invariant under the transformation law in \cite{Ma:lin} and
hence
behaves bosonically (i.e. with trivial braiding). We consider $\d$ as the
action of
the $V$ part of this element on $\CM$ by the action $\alpha$ defined by
differentiation\cite{Ma:fre},
followed by the product in $\Lambda$. This is shown diagrammatically in
Figure~1 (a). Part (b) recalls the
module-algebra property of the action $\alpha$\cite{Ma:fre} which comes out as
the braided-Leibniz
rule for the differentials ${\del\over\del x_i}=\alpha_{v^i}$ because the
coproduct $\Delta$ in the
degree one part $V_1$ of the vector algebra is just the linear one $v^i\tens
1+1\tens v^i$. Using these
facts about partial derivatives from \cite{Ma:fre} we can easily prove the
Leibniz rule for $\d$, which
we do in part (c). On the left is the braided tensor product\cite{Ma:bra} in
$\Lambda\und\tens \CM$, followed by
$\d$ in the box. We then use the braided-Leibniz rule from part (b) for the
first equality and functoriality
(rearrangement of braids) for the second, as well as associativity of the
products. In these diagrams we work in the braided category of $\tilde
A$-comodules
in which the braiding 
is given by $R$. But the
commutation relations of $\Lambda$ are also given
by $-R$ so we have
\[
\cdot\circ\Psi(\theta_{i}\tens\theta_{i_1}\cdots\theta_{i_p})
=(-1)^p\theta_{i_1}\cdots\theta_{i_p}\theta_i\]
which we use for the third equality. Note that we do not make use of the
coaddition on $\Lambda$, which would require
a different braiding (based on $-R'$) from the one we use here. Finally we use
functoriality and associativity again to
recognise the result. Conceptually, the element $\theta_i\tens{\del\over\del
x_i}$ is bosonic (invariant) and
hence the resulting derivation property is the usual $\Z_2$-graded one (albeit
coming out from the right in our present conventions) and not a braided one.
\endproof

This is the construction of the exterior differential calculus on a quantum or
braided vector space
coming out of braided geometry. The resulting R-matrix formulae
\eqn{dx}{ \d\vecx_1\d\vecx_2=-\d\vecx_2\d\vecx_1R,\quad
\vecx_1\d\vecx_2=\d\vecx_2\vecx_1 R}
here are essentially the same as in
\cite{WesZum:cov} but the difference is that we begin with our well-defined
partial
differential operators $\del\over\del x_i$ and define $\d$ from them in a
well-defined way, rather than arguing backwards by consistency requirements
within the axiomatic
framework of Woronowicz. In the braided approach the starting point is the
braided addition law\cite{Ma:poi}, which then defines partial
derivatives\cite{Ma:fre}, which in turn define $\d $ as above.

In nice cases, one will also have that $\d^2=0$. Using the above definition it
is clear that the essential requirement for this is the identity
\eqn{d2}{ \theta_1\theta_2\del_2\del_1=0}
which in turn is immediate at least when $PR'=f(PR)$ for some function $f$ such
that $f(-1)\ne 1$. For then
$\theta_1\theta_2\del_2\del_1=\theta_1\theta_2(PR')\del_2\del_1
=\theta_1\theta_2f(PR)\del_2\del_1=f(-1)\theta_1\theta_2\del_2\del_1.$
This includes the Hecke case where $R'\propto R$ but also the more general
construction\cite{Ma:poi} where $R'$ is built from the minimal polynomial
of $R$. In cases where $PR'$ is not given explicitly in terms of $PR$, we can
check (\ref{d2}) directly using the same strategy.  The same remarks apply
for the proof of covariance of the vector addition law in \cite{Ma:poi}.

Using these ingredients one can immediately write down a Laplacian
$L=\d\delta+\delta \d$ where $\delta=*\d*$ and develop a general theory of
wave-equations and electromagnetism on general braided spaces associated to
$R$-matrices. In our setting we consider the gauge potential
as $A\in \Omega^1$ and let $F=\d A$. The gauge theory is well-defined because
$\d^2=0$.
It is compatible with the Maxwell equations $\delta F=j$ because
$*^2\propto\id$. In other
words, the properties needed have been covered in our theory above and hold in
some generality.
More specific applications will be studied elsewhere, but see \cite{Mey:wav}
where
these concepts are developed directly for $q$-Minkowski with some interesting
results.

\section{Examples}

A simple example is the quantum plane $\C_q^n$ associated to the usual
$GL_q(n)$ $R$-matrix. In this
case our general formula for the $\eps$ tensor recovers the usual one as in
\cite{FRT:lie}. There is no
invariant metric so we do not have a Hodge $*$-operator. The differential
calculus recovers the one of
\cite{PusWor:twi}\cite{WesZum:cov}. This is clear already from the comparison
of the
corresponding partial derivatives in \cite{Ma:fre}.

On the other hand, many other important algebras in q-deformed physics are in
fact braided spaces with a coaddition
law, so at once amenable to our machinery. Note that once the additive braid
statistics $R$ is known, we do not
need to do any more work for the differential calculus: we just write down
(\ref{dx}) with {\em the same $R$} as used in the
braid statistics for the addition law on the quantum space.

For example, we know from \cite{Ma:add} that the quantum matrices
$A(R)=\{t^i{}_j\}$ have an addition law, at least when $R$ is Hecke. We can put
it into the form of a braided space by
\[ R\vect_1\vect_2=\vect_2\vect_1R\quad\Leftrightarrow t_It_J=t_Bt_A{\bf
R'}^A{}_I{}^B{}_J,\quad {\bf
R'}^I{}_J{}^K{}_L=R^{-1}{}^{j_0}{}_{i_0}{}^{l_0}{}_{k_0}
R^{i_1}{}_{j_1}{}^{k_1}{}_{l_1}\]
\[ \vect'_1\vect_2=R_{21}\vect_2\vect'_1R\quad\Leftrightarrow
t_It'_J=t'_Bt_A{\bf R}^A{}_I{}^B{}_J,\quad {\bf
R}^I{}_J{}^K{}_L=R{}^{l_0}{}_{k_0}{}^{j_0}{}_{i_0}
R^{i_1}{}_{j_1}{}^{k_1}{}_{l_1}\]
where $t_I=t^{i_0}{}_{i_1}$. We recover at once the vector algebra\cite{Ma:add}
and bicovariant differential calculus\cite{Sud:alg}
\eqn{dt}{ R\del_2\del_1=\del_1\del_2 R,\quad \d\vect_1\d\vect_2=-R_{21}\d
\vect_2\d\vect_1R,\quad \vect_1\d\vect_2=R_{21}\d \vect_2\vect_1R}
on $A(R)$ where $\del^I=\del^{i_1}{}_{i_0}={\del\over\del t^{i_0}{}_{i_1}}$ is
written as a matrix. This includes the usual results for $M_q(n)$ and
multiparametric $M_{p,q}(n)$ etc. That $d^2=0$ for
this class is known so we omit the direct check of (\ref{d2}) which is needed
in our constructive approach of the last section. It is very similar to the
proof for $\bar A(R)$ below. The construction is covariant under
$\widetilde{A}$ built from
$A(R)^{\rm cop}\tens A(R)$ as explained in\cite{Ma:add}, which corresponds to
bicovariance under $GL_q(n)$ etc. in the usual approach. The dilatonic
extension here is needed if one wants to go to quotient quantum groups rather
than working at the $GL_q$ level.
That $\bf R'$ obeys the QYBE etc. is easily checked and means that the theory
in Section~2 applies and gives us a $q$-epsilon tensor on $A(R)$ to go with
this differential calculus.

For a second class of examples, we have the algebras $\bar A(R)=\{x^i{}_j\}$
introduced in \cite{Ma:euc} as a variant of $A(R)$ and again with a braided
addition law when $R$ is Hecke. It forms a braided space with\cite{Ma:euc}
\[ R_{21}\vecx_1\vecx_2=\vecx_2\vecx_1R\quad\Leftrightarrow x_Ix_J=x_Bx_A{\bf
R'}^A{}_I{}^B{}_J,\quad {\bf
R'}^I{}_J{}^K{}_L=R^{-1}{}^{l_0}{}_{k_0}{}^{j_0}{}_{i_0}
R^{i_1}{}_{j_1}{}^{k_1}{}_{l_1}\]
\[ \vecx'_1\vecx_2=R\vecx_2\vecx'_1R\quad\Leftrightarrow x_Ix'_J=x'_Bx_A{\bf
R}^A{}_I{}^B{}_J,\quad {\bf R}^I{}_J{}^K{}_L=R{}^{j_0}{}_{i_0}{}^{l_0}{}_{k_0}
R^{i_1}{}_{j_1}{}^{k_1}{}_{l_1}\]
which gives at once the vector algebra and quantum differential calculus
\eqn{dxeuc}{ R\del_2\del_1=\del_1\del_2 R_{21},\quad
\d\vecx_1\d\vecx_2=-R\d\vecx_2\d\vecx_1R,\quad
\vecx_1\d\vecx_2=R\d\vecx_2\vecx_1R}
on $\bar A(R)$. Since this is a new differential calculus, we formally verify
(\ref{d2}) as
\align{&&\equad X=\trace \d \vecx_1 \d \vecx_2 \del_2\del_1=\trace \d \vecx_1
\d \vecx_2 R^{-1}R\del_2\del_1=-\trace R\d \vecx_2 \d \vecx_1
\del_1\del_2R_{21}=-\trace  \d \vecx_2 \d \vecx_1 \del_1\del_2(1+\lambda PR)\\
&&\equad 2X=-\lambda\trace \d \vecx_2 \d \vecx_1 R_{21}^{-1}R_{21}\del_1\del_2
PR=-\lambda\trace \d \vecx_2 \d \vecx_1 R_{21}^{-1}\del_2\del_1 P(1+\lambda
PR)\\
&&=\lambda\trace R_{21}\d \vecx_1\d\vecx_2\del_2\del_1
P-\lambda^2\trace\d\vecx_2\d\vecx_1\del_1\del_2=-2X-\lambda^2X}
where $\lambda=q-q^{-1}$ and $\trace=\trace_1\trace_2$ over the two sets of
matrix indices. Hence $X=0$ provided $4+\lambda^2\ne 0$, i.e. provided $q^2\ne
-1$. We just used the relations (\ref{dxeuc}) many times, cyclicity of the
trace and the Hecke assumption $R_{21}R=1+\lambda PR$.

It is also easy to see that $\bf R'$ obeys the QYBE etc. so that the theory
above applies and gives us a $q$-epsilon tensor on $\bar A(R)$ to go with this
differential calculus. Our constructions here are covariant under
$\widetilde{A}$ obtained from $A(R)\tens A(R)$. The simplest case with $R$ the
standard $4\times 4$ $SU_q(2)$ R-matrix gives $q$-Euclidean space\cite{Ma:euc}
and is studied in detail in Section~5.1. It is covariant under
$\widetilde{SU_q(2)\tens SU_q(2)}$, i.e. the $q$-Euclidean rotation group with
dilatonic extension.

For a third immediate class of examples, we know from \cite{Mey:new} that the
braided matrices $B(R)=\{u^i{}_j\}$ introduced in \cite{Ma:exa} also have a
braided addition law when $R$ is Hecke.  It appears in the form of a braided
space with\cite{Ma:exa}\cite{Mey:new}
\[ R_{21}\vecu_1R\vecu_2= \vecu_2 R_{21} \vecu_1 R\ \Leftrightarrow\
 u_Ju_L= u_Ku_I {\bf R'}^I{}_J{}^K{}_L,\quad {\bf
R'}^I{}_J{}^K{}_L=R^{-1}{}^{d}{}_{k_0}{}^{j_0}{}_{a}
R^{k_1}{}_{b}{}^{a}{}_{i_0}R^{i_1}{}_c{}^b{}_{l_1} {\widetilde
R}^c{}_{j_1}{}^{l_0}{}_d\]
\[ R^{-1}\vecu_1'R\vecu_2= \vecu_2 R_{21}\vecu_1'R\ \Leftrightarrow\
 u_J'u_L= u_Ku_I' {\bf R}^I{}_J{}^K{}_L,\quad {\bf
R}^I{}_J{}^K{}_L=R^{j_0}{}_a{}^d{}_{k_0}
R^{k_1}{}_b{}^a{}_{i_0}
R^{i_1}{}_c{}^b{}_{l_1} {\widetilde R}^c{}_{j_1}{}^{l_0}{}_d\]
where $\widetilde{R}$ is given by transposition in the second two indices,
inversion and transposition again. The first  multi-index R-matrix was
introduced by the author in \cite{Ma:exa} (as well as another one for
multiplicative braid statistics), while the second was introduced by Meyer in
\cite{Mey:new}. The algebra here is an important one and appears in other
contexts also as explained in \cite{Ma:skl}. For this algebra we have at once
the vector algebra and differential calculus
\eqn{du}{ R\del_2\widetilde{R}\del_1=\del_1\widetilde{R_{21}}\del_2
R_{21},\quad R^{-1}\d\vecu_1R\d\vecu_2=-\d\vecu_2 R_{21}\d\vecu_1R,\quad
R^{-1}\vecu_1R\d\vecu_2= \d\vecu_2 R_{21}\vecu_1R}
as recently studied by several authors\cite{Vla:coa}\cite{Isa:int} and
references therein. We would like to stress that these relations themselves are
just (and necessarily) the  same form as Meyer's additive braid statistics and
hence could not be considered as new. On the other hand, \cite{Vla:coa}
contains an interesting new result that $\Omega$ itself can have a braided
addition law in this and other cases. Meanwhile \cite{Isa:int} contains an
interesting observation about its braided-comodules.  For completeness, we need
to add from our constructive point of view the formal proof of (\ref{d2}) and
hence of $\d^2=0$ as
\align{&&\equad X=\trace \d \vecu_1 \d\vecu_2\del_2\del_1=\trace\d\vecu_1 R
\d\vecu_2\del_2 \widetilde{R}\del_1=\trace\d\vecu_1 R \d\vecu_2 R^{-1}R\del_2
\widetilde{R}\del_1\\
&&=-\trace R\d \vecu_2 R_{21}\d \vecu_1 \del_1
\widetilde{R_{21}}\del_2R_{21}=-\trace\d \vecu_2 R_{21}\d \vecu_1 \del_1
\widetilde{R_{21}}\del_2(1+\lambda PR)\\
&&\equad 2X=-\lambda\trace \d \vecu_2 R_{21}\d \vecu_1  R_{21}^{-1}R_{21}\del_1
 \widetilde{R_{21}}\del_2 PR=-\lambda\trace \d \vecu_2 R_{21}\d \vecu_1
R_{21}^{-1}\del_2 \widetilde{R}\del_1 P(1+\lambda PR)\\
&&=\lambda\trace R_{21}\d \vecu_1 R\d \vecu_2 \del_2 \widetilde{R}\del_1
P-\lambda^2\trace \d \vecu_2 R_{21}\d \vecu_1 \del_1
\widetilde{R_{21}}\del_2=-2X-\lambda^2X}
which implies $X=0$ provided $q^2\ne -1$. In fact, there is a mathematically
precise equivalence between this proof and the one for $A(R)$ and its variant
above for $\bar A(R)$, provided respectively by the theory of
transmutation\cite{Ma:exa} or twisting\cite{Ma:euc}. This expresses products of
the $\vecu$ in terms of products of the $\vect$ or $\vecx$ in a precise way and
in a corresponding way for the partial differentials $\del$.

It is also known that $\bf R'$ here obeys the QYBE\cite{Ma:sol}, while the
mixed relations involving $\bf R,R'$ are also easily checked in the same way
and reduce to the QYBE for $R$. Hence we are in the symmetrical situation
needed for our theory of the $\eps$ tensor. Moreover, our construction is
manifestly covariant under a quantum group $\widetilde{A}$ obtained from
$A(R)\dcross A(R)$, where $\dcross$ is the double cross product construction
from \cite{Ma:seq}. See also \cite[Sec. 4]{Ma:poi}.  The standard $4\times 4$
R-matrix gives $q$-Minkowski space studied in detail in Section~5.2. Here the
covariance is under $\widetilde{SU_q(2)\dcross SU_q(2)}$ which is the
$q$-Lorentz group of \cite{PodWor:def}\cite{CWSSW:lor} with dilatonic
extension.

In both cases here the dilatonic extension is needed for the braiding to be
given by our categorical constructions with the correct
normalisation\cite{Ma:poi}. One could try to leave it out by adjusting the
normalisations in (\ref{dx}) etc. by hand but in this case one can expect an
inconsistency at some other point where both the braiding and the determinant
or other non-quadratic relations of the background quantum group are needed. An
alternative way out is to allow the $q$-Lorentz group to be treated with
anyonic or $q$-statistics\cite{Ma:csta}. This will be explained elsewhere. We
note also that the $\bar A(R)$ and braided matrices $B(R)$ are related by a
twisting construction for comodule algebras\cite{Ma:euc} which becomes a
`quantum Wick rotation' in the Euclidean/Minkowski case. We know also from
their original definition in \cite{Ma:exa} that the braided matrices $B(R)$ are
strictly related to $A(R)$ by transmutation. This is already reflected in the
similarity of the proofs above and it would be interesting to formalise it
further as a theory of twisting and transmutation for differential calculi and
$\eps$ tensors.

We conclude with the simplest cases of the $\bar A(R)$ and $B(R)$ constructions
computed in detail using the symbol manipulation package REDUCE. This is needed
to determine the normalisations $d_m$ etc. concerning the projectors and Hodge
operators. Direct R-matrix methods like those above are not yet known for these
properties, but they are verified in both of these
examples as well as in similar ones based on other well-known R-matrices.

\subsection{q-Euclidean space}

For $q$-Euclidean space, we use the definition in \cite{Ma:euc} as twisting
$\bar M_q(2)$ of the usual $2\times 2$ quantum matrices. This is the simplest
example of the $\bar A(R)$ construction above. We have generators
$\vecx=\pmatrix{a&b\cr c&d}$ and relations
\ceqn{posneucl}{ba=qab,\quad ca=q^{-1}ac,\quad da=ad,\quad db=q^{-1}bd\quad
dc=qcd\\
bc=cb+(q-q^{-1})ad.}
This is actually isomorphic to the usual $M_q(2)$ by a permutation of the
generators, so one can
regard the following as results on this with its additive structure as
introduced in \cite{Ma:add}.

Vector algebra of derivatives is
\[ {\del\over\del d} {\del\over\del c}=q^{-1}  {\del\over\del c} {\del\over\del
d}, \quad {\del\over\del d} {\del\over\del b} = {\del\over\del b}
{\del\over\del d} q,\quad
{\del\over\del d} {\del\over\del a} = {\del\over\del a} {\del\over\del d}\]
\[ {\del\over\del c} {\del\over\del b}= {\del\over\del b} {\del\over\del c} +
(q-q^{-1} ) {\del\over\del a} {\del\over\del d},\quad
{\del\over\del c} {\del\over\del a} = {\del\over\del a} {\del\over\del c}
q,\quad {\del\over\del b} {\del\over\del a} =q^{-1}   {\del\over\del a}
{\del\over\del b}\]

The metric is \cite{Ma:euc}
\eqn{eucmetric}{\eta^{IJ}=\pmatrix{0&0&0&1\cr 0&0&-q&0\cr 0&-q^{-1}&0&0\cr
1&0&0&0}={\del\over\del x_I}{\del\over\del x_J}(ad-qcb).}
It has determinant $\det(\eta)=1$ and is already correctly normalised, so
$\lambda=1$ in (\ref{epsud}). Here
\[ x_Jx_I\eta^{IJ}=(1+q^{-2})(ad-qcb)\]
is a natural `radius function' according to the construction in \cite{Ma:euc}.

The algebra of forms is
\[ \d a \d a=0,\quad \d b\d b=0,\quad \d c\d c=0,\quad \d d\d d=0\]
\[ \d b  \d a =-q^{-1}   \d a  \d b ,\quad \d c  \d a  =- \d a  \d c  q,\quad
\d d   \d b  =- \d b  \d d   q\]
\[ \d c  \d b  =- \d b  \d c ,\quad \d d   \d c =-q^{-1}   \d c  \d d  ,\quad
\d d   \d a =- (q-q^{-1} ) \d b  \d c  - \d a  \d d  \]

We have the $q$-epsilon tensor as:
\[\eps_{abcd}=-\eps_{acbd}=\eps_{adbc}=-\eps_{adcb}
=\eps_{bcad}=-\eps_{bcda}=1\]
\[-\eps_{cbad}=\eps_{cbda}=-\eps_{dabc}=\eps_{dacb}
=-\eps_{dbca}=\eps_{dcba}=1\]
\[ \eps_{acdb}=-\eps_{cdba}=-\eps_{dcab}=q,\quad
-\eps_{abdc}=-\eps_{bacd}=\eps_{bdca}=\eps_{dbac}=q^{-1}\]
\[ -\eps_{cadb}=\eps_{cdab}=q^2,\quad  \eps_{bacd}=-\eps_{bdac}=q^{-2}\]
\[ -\eps_{adad}=\eps_{dada}=(q-q^{-1})\]

The resulting raw (un-normalised) antisymmetriser projectors $P$ have
associated constants
\[ d_0=2q^4[2]^2[3],\quad d_1=-2q^2[3],\quad d_2=q^2[2]^2,\quad
d_3=-2q^2[3],\quad d_4=2q^4[2]^2[3]\]
where $[m]\equiv[m,q^{-2}]={1-q^{-2m}\over 1-q^{-2}}$. In each case, the
projections are on the space of totally $R$-antisymmetric tensors and have the
same ranks as classically.

The Hodge $*$-operator for this metric is:
\[ (\d a\d b\d c\d d)^*=1,\quad (\d a\d b\d c)^*=\d a,\quad (\d a\d b\d d)^*=\d
b\]
\[ (\d a\d c\d d)^*=-\d c,\quad (\d b \d c\d d)^*=-\d a\]
\[ (\d a\d b)^*=-q[2]\d a\d b,\quad (\d a\d c)^*=q[2]\d a\d c,\quad (\d a\d
d)^*=2\d b\d c-(q-q^{-1})\d a\d d\]
\[ (\d b\d c)^*=2\d a\d d+(q-q^{-1})\d b \d c,\quad (\d b\d d)^*=q[2]\d b\d
d,\quad (\d c\d d)^*=-q[2]\d c\d d\]
\[ (\d a)^*=2q^2[3]\d a\d b\d c,\quad (\d b)^*=2q^2[3]\d a\d b\d d,\quad (\d
c)^*=-2q^2[3]\d a\d c\d d,\quad (\d d)^*=-2q^2[3]\d b\d c\d d\]
\[ 1^*=2q^4[2]^2[3]\]
One can check that the square of this Hodge * operator is
\[ *^2=(-1)^mP=(-1)^md_m\]
on forms of degree $m$. The Hodge $*$-operator on 2-forms is a $6\times 6$
matrix with eigenvalues $\pm q[2]$ with multiplicity $3$. The self-dual and
antiself dual 2-forms with respect to it are characterised by
\[  F^*=q[2]F,\quad {\rm (self-dual\ form)}; \quad  F^*=-2[q]F,\quad {\rm
(anti-self-dual\ form)}.\]
Of course, one may adjust the normalisation of $*$ to have the more usual
limiting form.

Note that our computations here have been for a matrix basis where the metric
$\eta$ has
the signature (2,2) in the $q=1$ limit. The $\eps$ tensor and value of $*^2$
are as one would expect for
this after bearing in mind the ordering of the indices in our conventions
(there is a reversal in (\ref{hodge})). There is a complex transformation with
real determinant which maps the matrix basis to the usual space-time basis
$t,x,y,z$ with Euclidean signature, so in this basis we still have $*^2$
positive. Again this is the right classical result for our index
conventions. The same remarks apply for the quantum case with $q$ real. The
noncommutative matrix generators transform to self-adjoint or `real' ones by a
complex linear transformation\cite{Ma:euc}. The $\eps$ computed in the new
basis is not just tensorially related to
the one in the matrix basis because the top form $\d t \d x \d y \d z$ is
different from $\d a \d b \d c \d d$ that we differentiated before. But these
top forms are proportional, up to a real constant and $\eps$ transforms
tensorially up to this.

By way of contrast, we include also the second Hodge $\circ$-operator:

\[ (\d a\d b\d c\d d)^\circ=1,\quad (\d a\d b\d c)^\circ=-\d a,\quad (\d a\d
b\d d)^\circ=-\d b\]
\[ (\d a\d c\d d)^\circ=\d c,\quad (\d b \d c\d d)^\circ=\d d\]
\[ (\d a\d b)^\circ=-q \d a\d b,\quad (\d a\d c)^\circ=q^{-1}\d a\d c,\quad (\d
a\d d)^\circ=\d b\d c-(q-q^{-1})\d a\d d\]
\[ (\d b\d c)^\circ=\d a\d d,\quad (\d b\d d)^\circ=q^{-1}\d b\d d,\quad (\d
c\d d)^\circ=-q\d c\d d\]
\[ (\d a)^\circ=-\d a\d b\d c,\quad (\d b)^\circ=-\d a\d b\d d,\quad (\d
c)^\circ=\d a\d c\d d,\quad (\d d)^\circ=\d b\d c\d d\]
On two-forms it has eigenvalues
$q^{-1},-q$ each with multiplicity 3. Hence with respect to $\circ$ we have
\[ F^\circ=q^{-1}F,\quad {\rm (self-dual\ form)}; \quad  F^\circ =-qF,\quad
{\rm (anti-self-dual\ form)}\]

\subsection{q-Minkowski space}

We use for $q$-Minkowski space the $2\times 2$ braided-hermitian matrices
introduced in \cite{Ma:exa}. It is the simplest example of the $B(R)$
construction above.
The covector algebra of position co-ordinates
$\vecu=\pmatrix{a&b\cr c&d}$ is:
\ceqn{posn}{ba=q^2ab,\quad ca=q^{-2}ac,\quad d a=ad,\qquad
bc=cb+(1-q^{-2})a(d-a)\\
d b=bd+(1-q^{-2})ab,\quad cd=d c+(1-q^{-2})ca}
This maps onto a braided tensor product of two copies of the quantum plane and
is thereby compatible with the approach of \cite{CWSSW:lor}\cite{OSWZ:def}
also. The additive
structure we need is from \cite{Mey:new}.

The vector algebra of differentiation operators is:
\[ {\del\over\del d}{\del\over\del c}=q^{-2}{\del\over\del c}{\del\over\del
d},\quad
{\del\over\del d}{\del\over\del b}= {\del\over\del b}{\del\over\del d}q^2\]
\[{\del\over\del d}{\del\over\del a}= {\del\over\del a}{\del\over\del d},\quad
{\del\over\del b}{\del\over\del a}={\del\over\del a}{\del\over\del
b}+{\del\over\del b}{\del\over\del d}(q^2-1)\]
\[{\del\over\del c}{\del\over\del a}={\del\over\del a}{\del\over\del
c}+{\del\over\del c}{\del\over\del d}(q^{-2}-1),\quad
{\del\over\del c}{\del\over\del b}={\del\over\del b}{\del\over\del
c}+{\del\over\del d}{\del\over\del d}(q^{-2}-1)+{\del\over\del a}{\del\over\del
d}(q^2-1)\]
The metric is \cite{Mey:new}:
\eqn{minkg}{\eta^{IJ}={\del\over\del u_I}{\del\over\del
u_J}(ad-q^2cb)=\pmatrix{q^{-2}-1&0&0&1\cr 0&0&-q^2&0\cr 0&-1&0&0\cr
1&0&0&0}.}
It has $\det(\eta)=q^{2}=\lambda$ as the required normalisation constant in
(\ref{epsud}). We get back the `radius function' from the metric as
\[ u_Ju_I\eta^{IJ}=(1+q^{-2})(ad-q^2cb).\]

The algebra of forms is
\[ \d c \d c=0,\quad \d a \d a=0,\quad \d b \d b=0,\quad \d b \d a=-\d a \d b\]
\[ \d c \d a=-\d a \d c,\quad \d c \d b=-\d b \d c,\quad\d d\d d=\d b \d
c(1-q^{-2})\]
\[\d d \d c=-\d c\d d q^{-2}+\d a \d c(1-q^{-2}),\quad\d d \d b=- \d b\d d q^2
- \d a \d b(q^2 -1)\]
\[dd \d a=- \d b \d c (q^2 -1)- \d a\d d\]
The dimensions in each degree are the usual ones: 1:4:6:4:1 and we can take a
basis $\d x_{i_1}\cdots \d x_{i_m}$ with $i_1<i_2\cdots<i_m$ with top form $\d
a \d b \d c\d d$.

We have the $q$-epsilon tensor as:

\[\eps_{addd}=-\eps_{bdcd}=-\eps_{dadd}
=\eps_{dbdc}=\eps_{ddad}=-\eps_{ddda}=1-q^{-2}\]
\[ -\eps_{adad}=-\eps_{cdbd}=\eps_{dada}=\eps_{dcdb}=q^2-1\]
\[\eps_{abcd}=-\eps_{acbd}=\eps_{adbc}=-\eps_{adcb}
=-\eps_{bacd}=\eps_{bcad}=-\eps_{bcda}=\eps_{cabd}=1\]
\[-\eps_{cbad}=\eps_{cbda}=-\eps_{dabc}=\eps_{dacb}
=\eps_{dbac}=-\eps_{dbca}=-\eps_{dcab}=\eps_{dcba}=1\]
\[ \eps_{acdb}=-\eps_{cadb}=\eps_{cdab}=-\eps_{cdba}=q^2\]
\[ -\eps_{abdc}=\eps_{badc}=-\eps_{bdac}=\eps_{bdca}=q^{-2}.\]

The resulting raw (un-normalised) antisymmetriser projectors have associated
constants
\[ d_0=2q^4[2]^2[3],\quad d_1=-2q^2[3],\quad d_2=q^2[2]^2,\quad
d_3=-2q^2[3],\quad d_4=2q^4[2]^2[3]\]
as in the Euclidean case. The corresponding projections are on the space of
totally $R$-antisymmetric
tensors and have the same ranks as classically.

The Hodge $*$-operator for this metric is:
\[ (\d a\d b\d c\d d)^*=q^{-2},\quad (\d a\d b\d c)^*=q^{-2}\d a,\quad (\d a\d
b\d d)^*=q^{-2}\d b\]
\[ (\d a\d c\d d)^*=-\d c,\quad (\d b \d c\d d)^*=q^{-2}(1-q^{-2})\d a-q^{-2}\d
d\]
\[ (\d a\d b)^*=-[2]\d a\d b,\quad (\d a\d c)^*=[2]\d a\d c,\quad (\d a\d
d)^*=2\d b\d c-(1-q^{-2})\d a\d d\]
\[ (\d b\d c)^*=2q^{-2}\d a\d d+(1-q^{-2})\d b \d c,\quad (\d b\d d)^*=[2](\d
b\d d+2(1-q^{-2})\d a\d b),\quad (\d c\d d)^*=-[2]\d c\d d\]
\[ (\d a)^*=2q^2[3]\d a\d b\d c,\quad (\d b)^*=2q^2[3]\d a\d b\d d,\quad \]
\[(\d c)^*=-2q^2[3]\d a\d c\d d,\quad (\d d)^*=-2q^2[3](\d b\d c\d d
-(1-q^{-2})\d a \d b \d c),\quad 1^*=2q^4[2]^2[3]\]
One can check that the square of this Hodge * operator is
\[ *^2=(-1)^mq^{-2}P=(-1)^mq^{-2}d_m\]
on forms of degree $m$. The Hodge $*$-operator on  2-forms has eigenvalues $\pm
[2]$ with multiplicity $3$. The self-dual and antiself dual 2-forms with
respect to it are characterised by
\[ F^*=[2]F,\quad {\rm (self-dual\ form)}; \quad  F^*=-[2]F,\quad {\rm
(anti-self-dual\ form)}.\]
As before, one can adjust the normalisation of $*$ to have the more usual limit
when $q=1$.

Also, the same remarks apply as in the Euclidean case to the effect that there
is a natural $*$-structure and a complex transformation from our matrix basis
to self-adjoint or `real' space-time bases $t,x,y,z$\cite{Ma:mec}. This time
the top form, $\eps$ and $*$ change by an imaginary factor. This again brings
our results here in line with the classical situation for our indexing
conventions.

Finally, by way of contrast our alternative Hodge $\circ$-operator is
 \[ (\d a\d b\d c\d d)^\circ=q^{-2},\quad (\d a \d b\d d)^\circ=- \d b q^{-2}\]
\[ (\d a \d c\d d)^\circ=\d c,\quad (\d a\d d\d d)^\circ=- \d a
q^{-2}(1-q^{-2}),\quad
(\d b \d c\d d)^\circ=ddq^{-2} - \d aq^{-2}(1-q^{-2})\]
\[ (\d a \d b)^\circ= - \d a \d b, \quad (\d a\d c)^\circ=\d a \d c
q^{-2},\quad (\d a\d d)^\circ=\d b \d c+ \d a\d d (q^{-2}-1), \quad (\d b \d
a)^\circ=\d a \d b\]
\[ (\d b \d c)^\circ=\d a\d d q^{-2},\quad (\d b\d d)^\circ=\d b\d d q^{-2} +
\d a \d b(1-q^{-4}),\quad (\d c\d a)^\circ=- \d a \d c q^{-2}\]
\[ (\d a)^\circ= - \d a \d b \d c,\quad (\d b)^\circ= - \d a \d b\d d,\quad
(\circ \d c)=\d a \d c\d d q^{-2},\quad (\d d)^\circ=\d b \d c\d d - \d a \d b
\d c(1-q^{-2})\]
On two-forms it has eigenvalues
$q^{-1},-1$ each with multiplicity 3. Hence  with respect to $\circ$ we have
\[ F^\circ =q^{-1}F,\quad {\rm (self-dual\ form)}; \quad  F^\circ =-F,\quad
{\rm (anti-self-dual\ form)}\]


\end{document}